\documentclass[letter,nofootinbib,12pt,superscriptaddress]{revtex4-2}

\usepackage[colorlinks,citecolor=blue]{hyperref}
\usepackage{amsmath}
\usepackage{mathrsfs}
\usepackage{bbm}
\usepackage{amsfonts}
\usepackage{amssymb}
\usepackage{latexsym}
\usepackage{graphicx}
\usepackage[english]{babel}
\usepackage{multirow}
\usepackage{float}
\usepackage{url}
\usepackage{slashed}
\usepackage[compat=1.1.0]{tikz-feynman}
\usepackage{orcidlink}
\begin{document}

\title{Dynamic Lasing of Axion Clusters}
 
\author{Liang Chen
\orcidlink{0000-0002-0224-7598}}
\email{bqipd@protonmail.com}
\affiliation{Research Center of Applied Physics and Photoelectric Information, Chizhou University, Chizhou, Anhui, 247000, China}

\author{Thomas W. Kephart
\orcidlink{0000-0001-6414-9590}}
\email{tom.kephart@gmail.com}
\affiliation{Department of Physics and Astronomy, Vanderbilt University, Nashville, TN 37235, USA}

\date{\today}

\begin{abstract}
\noindent 
We examine high-density axion clusters under gravitational compression. These are transient events in which the majority of axions are rapidly converted into photons, with some configurations producing photon signals with distinctive and characteristic patterns. We estimated the mass of the remnant objects and note that some could be black holes while in some cases it may be possible to identify the emitted photons with a robust class of fast radio bursts.

\end{abstract}

\maketitle
\newpage

\section{Introduction}
\label{sec-intro}

One reason the QCD axion is considered to be one of the most promising candidates for the particle that constitutes dark matter in the Universe, is that its existence also provides a solution to the strong CP problem. Due to the spontaneous breaking of the Peccei-Quinn \cite{Peccei:1977hh,Peccei:1977ur} symmetry, the axion $a$ appears as a spin-0 pseudo-Goldstone boson~\cite{Weinberg:1977ma}, while the explicit breaking of the global axial $U_A(1)$ symmetry via the chiral anomaly of QCD gives the axion its mass. 
In models where the axion arises as a pseudo-Goldstone boson, the interaction Lagrangian at energy scales below the symmetry breaking scale, characterized by the decay constant $f_a$, is given by:
$f_a^{-1} J^\mu \partial_\mu a$,
where $a$ is the axion field and $J^\mu$ denotes the Noether current of the spontaneously broken global symmetry. Thus the axion's interactions with Standard Model particles are suppressed by the decay constant $f_a$. Note that the axion models with $f_a$ assumed at the electroweak symmetry breaking scale $v_\text{weak}\sim250$~GeV have been ruled out by previous experiments~\cite{Barshay:1981ky, Barroso:1981ta}, which favor benchmark models such as KSVZ~\cite{Kim:1979if, Shifman:1979if} and
DFSZ~\cite{Dine:1981rt, Zhitnitsky:1980tq}, predicting ``invisible'' axions evading almost all current experimental limits with $f_a\gg v_\text{weak}$.
By including $O(\alpha)$ QED corrections and next-to-next-to-leading order corrections in chiral perturbation theory, Ref.~\cite{Gorghetto:2018ocs} finds
\begin{flalign}\label{}
m_a f_a = 5.691 \times 10^9 \text{ meV GeV}.
\end{flalign}
Inherited from axion's mixing with $\pi^0$,
its interactions generally scale approximately with those of the $\pi^0$, i.e., $m_{\pi}f_{\pi} \approx m_a f_a$.
The axion's two-photon vertex reads
\begin{flalign}\label{2-photon}
{\cal L}_{a\gamma\gamma} = 
{\alpha K \over 8\pi f_a} a F_{\mu\nu} \tilde F^{\mu\nu}~,
\end{flalign}
where $\alpha$ is the fine structure constants of electromagnetic interactions with $F_{\mu\nu}$ and $\tilde{F}_{\mu\nu}$ as the field strength and the corresponding dual field, respectively, and $K$ is a model dependent coefficient of ${\cal O}(1)$. 
It enables the primary search strategy based on axion-photon conversion\cite{Sikivie:1983ip}, an effect that may also have astrophysical significance.
The decay process $a\to \gamma\gamma$ gives the axion lifetime,
\begin{flalign}\label{lifetime1}
\tau_a = {256\pi^3 \over K^2\alpha^2 m_a} \left({  f_a \over m_a }\right)^2\,,
\end{flalign}
and the axion's lightness results in a $\tau_a$ longer than the age of the universe if $m_a\lesssim$ a few eV. We set the model-dependent parameter $K$ to be 1 by default, though it could range from approximately 1 to 10 \cite{Cheng:1995fd}.

Our interest here is in the possible detection of dense axion clouds (clusters) decaying via stimulated emission to photons. 
Tkachev\cite{Tkachev:1986tr} examined the possibility that the growing axion density at the center of
a gravitational well could lead to to a coherent cosmic maser source, driven by the stimulated
$a\rightarrow\gamma\gamma$ process during the galaxy formation era, and discussed the scenario in which gravitational condensation leads to the formation of spherically symmetric configurations with cosmological strings acting as the seeds\cite{Tkachev:1987cd}.
Kephart and Weiler initially considered solely the mechanism of spontaneous decay of the axion\cite{Kephart:1986vc}, later providing detailed calculations of the stimulated emission rate for axion clusters\cite{Kephart:1994uy,Kephart:1999ti}. In these early studies, gravity merely serves as a mechanism to confine axions within a stationary system. 
Braaten et al.\cite{Braaten:2016dlp} calculated the rate at which axions are lost from axion stars via axions annihilating into photons but they considered the spontaneous emission only. (For a review of the physics of axion stars see \cite{Braaten:2019knj}, where stable nonlasing axion configurations have also been studied.) Levkov et al.\cite{Levkov:2020txo} offered a quasistationary approach to parametric resonance in a finite volume for nonrelativistic axions, explicitly addressing that this type of resonance can not be blocked by gravitational and self-interaction. They found that the strength of axion-photon couplings for conventional QCD axions is insufficient for axion stars to develop parametric resonance.
Hertzberg and Schiappacasse\cite{Hertzberg:2018zte} came to a similar conclusion, indicating that conventional QCD axion clumps would not lead to photon resonance. 

There are several differences between our work in this paper and the aforementioned studies. While parametric resonance is rooted in classical physics, stimulated emission is a purely quantum phenomenon. They are not entirely equivalent or interchangeable. Therefore, the conclusions from these studies, which suggest that conventional QCD axion clusters cannot induce parametric resonance, do not rule out the possibility of stimulated emission.
Moreover, the studies in Refs.~\cite{Hertzberg:2018zte,Braaten:2016dlp}  investigate the parametric resonance of stable axion stars, or, at most, the collapse of critical axion stars\cite{Levkov:2020txo}.

There are several unclear aspects in the investigations of parametric resonance involving ``stable'' axion clumps. First, by definition, stable axion clumps should not lead to photon resonance, yet current theories on stable axion configurations seem to overlook this, allowing for the possibility of clumps that could induce photonic resonance. Second, the descriptions of the evolution of stable axion clumps during and after photonic resonance are lacking in detail. If these stable axion clusters or clumps remain intact for an extended period during and after parametric resonance, then the predicted photon emissions will need to be mild and occur steadily over time. However, these remain open questions so far in literature. 

Since there is no guarantee that all axion stars or clumps are stable or subcritical, it is likely that stable ones represent only a portion of the entire axion cluster population. Our research focuses on stimulated emission from unstable axion clusters, particularly those undergoing self-gravitational compression. We believe it is more logical to begin investigations with unstable axions rather than ``stable'' ones, given the arguments we have presented. If any of these axion clusters lased, the resulting events would be expected to be transient, characterized by the rapid conversion of axions into photons. Certain configurations could produce photon signals with distinct and recognizable patterns.

A stationary lasing model of axion clusters was proposed  in Refs~\cite{Kephart:1994uy,Kephart:1999ti}, along with minimum density $\rho_1$ required for lasing. We describe this model as stationary because it does not take into account the self-gravitational compression of axion clusters, despite that such compression occurs. Although stimulated emission occurs rapidly in dense axion clusters, it is a reasonable approximation for this model to neglect gravitational compression. Nevertheless this approximation makes the stationary model inadequate for describing dynamic scenarios. For example, dilute axion clumps cannot lase according to the stationary model, but they undergo gravitational compression into high-density axion clusters, potentially triggering stimulated emissions in the process.
On the other hand, the stimulated emission of axions can influence the dynamics of gravitational compression by reducing the mass of the axion clump, thereby prolonging the collapse. These effects are now properly incorporated into a dynamic lasing model that we propose here. 
Related works that build upon Refs.~\cite{Kephart:1994uy,Kephart:1999ti} include Refs.~\cite{Chen:2020ufn,Chen:2020yvx,Chen:2020eer,Chen:2023bne}. A more detailed review of this topic can be found in Ref.~\cite{Chen:2023jki}.

This paper is organized as follows. We begin by briefly reviewing the stimulated emission process and the self-gravitational compression of an axion clump. Next, we derive the evolution equations by integrating the Boltzmann equation. We then utilize these evolution equations to present the results of several examples. Finally, we discuss the possible connection between the photonic bursts from axion clumps and fast radio bursts (FRBs), as well as the remnant objects resulting from the collapse of axions.

\section{Stimulated emission from an axion cluster}\label{SecEq}
In this section, let us begin by describing the formalism of the evolution of the axion- photon system. The Boltzmann equation which governs the rate of change of the photon number density $n_{\lambda}$ of helicities $\lambda=\pm 1$ reads\cite{Kephart:1986vc,Kephart:1994uy}, 
\begin{flalign}\label{BoltzmannEq}
\frac{dn_{\lambda}}{dt}=&\int dX^{(3)}_\text{LIPS}[f_a(1+f_{1\lambda})(1+f_{2\lambda})-f_{1\lambda}f_{2\lambda}(1+f_a)]|M(a\rightarrow \gamma \gamma)|^2 ~,
\\\nonumber  \text{with }
\int dX^{(3)}_\text{LIPS} =&  \int {d^3p\over (2\pi)^3 2p^0} \int {d^3k_1\over (2\pi)^3 2k_1^0} \int {d^3k_2\over (2\pi)^3 2k_2^0} 
 (2\pi)^4 \delta^{(4)}(p-k_1-k_2) ~.
\end{flalign}
where $p$, $k_i$ and $f_a(p,t)$, $f_{i\lambda}(k_i,t)$  are the momenta and occupation numbers of the axion and the photons ($i=1,2$ since an axion decays into two photons) respectively; $\delta^{(4)}$ ensures conservation of energy-momentum during the evolution process. For any particle species $i=\gamma$ or $a$ with the occupation number $f_i$, the number density $n_i$ and the total number $N_i$ of these particles are
\begin{flalign}
n_i= \int \frac{d^3p}{8\pi^3}f_i\,, \quad\text{and}\quad N_i = \int_V d^3r\,\, n_i ,
\end{flalign}
respectively. The photon field is parity
symmetric, $f_+(\vec{k})=f_-(\vec{k})={1\over2}f_\gamma(\vec{k})$ with $f_\gamma$ being the total photon occupation number after helicities being summed. Our discussion emphasizes axions and photons in the system, neglecting photon attenuation from non-axionic matter or frequency shifting from scattering in the local environment, etc. $M(a\rightarrow \gamma \gamma)$ denotes the amplitude of an axion decay into a photon pair, which can be derived from the axion-photon interaction in Eq.~\eqref{2-photon} and is linked to the lifetime of spontaneous axion decay via
\begin{flalign}\label{lifetime2}
\tau_a = {1 \over 32\pi m_a} \sum_{\lambda=\pm} 
|M(a\rightarrow \gamma \gamma)|^2
~.
\end{flalign} 

We aim at obtaining the photon number density $n_\lambda$ from a given axion occupation number $f_a$ by integrating the Boltzmann evolution equation~\eqref{BoltzmannEq}. If the momentum dependence of the axions have a bias aligning them towards certain direction, then the photons from spontaneous decay of these axions would not inherit this directional bias. This is because in the process $a\rightarrow \gamma \gamma$, the photon momenta $k_1, k_2$ (which are back-to-back in the axion rest frame) can point in any direction, regardless of how the initial axion momentum $p$ was oriented. Even if the momentum directions of photons produced by stimulated decay align with the incident photon, due to a lack of directional bias in the momenta of initial photons produced by spontaneous decay, the momentum distribution of total photons still does not possess a preferred direction, which paves the way for an ansatz that the occupation number of photons $f_\lambda$ is spherically symmetrical, at least in momentum space. Note that if the initial photon occupation number is extremely high and exhibits an inherent directional bias, this analysis might not hold. However, this is not the case for the following discussions, where the initial photon distribution is very dilute.

Consider now the gravitational contraction of an axion cluster. Pressureless free collapse of the cluster allows the axion occupation number $f_a$ to have a uniform radial dependence $\Theta(R-r)$, as uniform density is maintained during pressureless spherical gravitational collapse  \cite{Oppenheimer:1939ue,Misner:1973prb}. 
As stated in Ref.~\cite{Kolb:1993zz}, it is reasonable to neglect the spatial gradients of the axion field in the equations of motion for temperatures below the QCD scale where axions can be treated
as pressureless, cold dust.
We will study radial movement of the axions without rotation, which coincides with the irrotational velocity field of self-gravitating Bose-Einstein axion condensate in analyses such as Refs.~\cite{Chavanis:2011zi,Chavanis:2016dab}. The relation between initial radial coordinate $r_0$ and velocity $v_0$ of an axion of mass $m_a$, radial coordinate $r$, velocity $v$ can be easily found in the nonrelativistic limit to be
\begin{flalign}\label{eom1}
 - G  m_0 {m_a \over r^2}=& m_a \ddot r
~\Rightarrow~
v = - \sqrt{ 2 G m_0 ( {1\over r} - {1\over r_0} ) + v_0^2 } ~,
\end{flalign}
where $m_0$ is the total mass of axions inside radius $r_0$, and $G$ is gravitational constant. As will be shown later, applications of non-relativistic mechanics with Newtonian gravity are justified, as both the terminal speeds of axions and the radii of axion cluster are well below relativistic limits. Field theoretic studies \cite{Chavanis:2011zi,Chavanis:2016dab,Hertzberg:2018zte,Levkov:2020txo} of this topic also handle axion clumps in the non-relativistic regime.

These observations allow one to make an ansatz regarding the axion and photon occupation numbers. Taking the axion occupation number proposed in stationary model\cite{Kephart:1994uy} as the prototype, we approximate the phase space distribution of axion as
\begin{flalign}\label{DistA}
f_a(\vec{p}, r, t) =& f_a(t) {m_a^3 \over p^2\sin\theta_p}  \delta[ p - P(r) ]\, \delta[\theta_p-(\pi-\theta)]  \delta[\phi_p-(\pi-\phi)] \Theta(R-r) ~.
\end{flalign}
In the process of free spherical compression, the two $\delta$-functions, $\delta[\theta_p - (\pi - \theta)]$ and $\delta[\phi_p - (\pi - \phi)]$, enforce the alignment of the axion momentum with the cluster's radial direction. Meanwhile, the $\delta$-function $\delta[ p - P(r)]$ ensures that the magnitude of the axion momentum corresponds to $P(r)$, as derived from Eq.~\eqref{eom1}. The prefactor ${m_a^3 \over p^2 \sin \theta_p}$ guarantees that the occupation number has the correct dimensionality. The step function $\Theta(R-r)$ follows the lasing model in Ref.~\cite{Kephart:1994uy} and demands the lasing phenomenon only take place inside an the axion cluster radius $R$. Since we consider gravitational compression, $R$ is allowed to change in the process. The temporal coefficient $f_a(t)$ encodes the variation of axion phase space density.

Suppose the 4-momentum vector of an axion is $p^\mu=(p^0,\vec{p})$ with $p\equiv|\vec{p}|=m_a c \beta$. It decays into two photons with 4-momentum vector $k^\mu=(k^0,\vec{k})$ and $p^\mu-k^\mu=(p^0-k^0,\vec{p}-\vec{k})$, respectively. Depending on the angle between the directions of the outgoing photon momentum $\vec{k}$ and the initial axion momentum $\vec{p}$, conservation of energy and momentum requires the magnitude of momenta of the outgoing photons stay within a certain range\cite{Kephart:1994uy}, $k_-\leq k\equiv|\vec{k}|=k^0 \leq k_+$, where
\begin{flalign}\nonumber
k_\pm =& { p^0 \pm p \over2} 
=  { m_a  ( 1 \pm \beta ) \over2 \sqrt{1-\beta^2} }~,~~
\beta={1\over c}\sqrt{ 2 G m_0 ( {1\over R} - {1\over R_0} ) +   v_0^2 } ~.
\end{flalign}
At any point $r$ in the axion cluster, we assume that the momenta of photons can take any values in the allowed range with equal probability. Thus the phase space distribution of photon can be characterized by
\begin{flalign}\label{DistP}
 f_\lambda(\vec{k}, r, t) =& f_\lambda(t)  \Theta(k_+-k) \Theta(k-k_-) \Theta(R-r) ~.
\end{flalign}
The temporal coefficient $f_\lambda(t)$  encodes the variation of photon phase space density.

The number density of axions can be written as $n_a(r, t) = n_a(t) \Theta(R-r)$ because of the factorization of axion occupation number in Eq.~\eqref{DistA}. Integrating axion occupation number $f_a(\vec{p}, r, t)$ over momentum space gives the axion number density $n_a(r, t)$ and leads to a relation between temporal coefficients $f_a(t)$ and $n_a(t)$,
\begin{flalign}\nonumber
 n_a(r, t) = n_a(t) \Theta(R-r) 
=& \int f_a(\vec{p}, r, t) {d^3 p\over(2\pi)^3}
~\Rightarrow~ f_a(t) = { 8\pi^3 n_a(t) \over m_a^3 } ~.
\end{flalign}
Similarly, integrating photon occupation number $f_\lambda(\vec{k}, r, t)$ over momentum space gives photon number density $n_\lambda(r, t)$ and leads to a relation between temporal coefficients $f_\lambda(t)$ and $n_\lambda(t)$,
\begin{flalign}\nonumber
& n_\lambda(r, t) = n_\lambda(t) \Theta(R-r) 
= \int f_\lambda(\vec{p}, r, t) {d^3 k \over (2\pi)^3}
~\Rightarrow~
f_\lambda(t) = { 8\pi^2 n_\lambda(t) \over   m_a^3 \beta } ~.
\end{flalign}

\section{Evolution equations}
Integrating the Boltzmann equation \eqref{BoltzmannEq} over momentum spaces of $\vec{p}$, $\vec{k}$ and $\vec{k}_1$, and replacing coefficients $f_{a(\lambda)}(t)$ with $n_{a(\lambda)}(t)$, we obtain
\begin{flalign}\label{d-lambda-dt}
{d n_\lambda(t)   \over dt}
=&\Gamma_{a} \{     n_a(t) 
[ 1 +   { 16\pi^2 n_\lambda(t) \over   m_a^3 \beta } ]
  -  { 16\pi^2  [n_\lambda(t)]^2 \over m_a^3  }  (  {2 \beta\over3}  + 1 ) \} ~.
\end{flalign}
For more specific integration procedures, see Refs.~\cite{Chen:2020ufn,Chen:2020yvx,Chen:2020eer} where integration of this sort has been carried out in a variety of contexts. The left-hand side of Eq. \eqref{d-lambda-dt} represents the rate of change of the photon number density. The right-hand side encompasses the various contributions to this change, including spontaneous axion decay, stimulated decay of axions induced by photons, and photon back-scattering into axions. These three mechanisms are built into the Boltzmann equation \eqref{BoltzmannEq} at the microscopic level of individual particle interactions $a\leftrightarrow \gamma \gamma$. However, while accounting for these interactions locally, the Boltzmann equation itself is not concerned with the exterior environment where these interactions occur at the macroscopic level, in particular, the size of an axion cluster. 

Photons with speed $c$ are constantly leaving the surface of a cluster of radius $R$ before being able to stimulate or scatter back  into axions. This surface loss rate of photons is
\begin{flalign}\nonumber 
\left( {d n_\lambda\over dt}  \right)_{\text{surface loss}} & = - {3c\over2R} n_\lambda ~.
\end{flalign}
In addition, the radius $R$ is shrinking during gravitational compression, which inherently would increase the number density of photons and axions. This enhancement rate of photon is
\begin{flalign}\nonumber
\left( {d n_\lambda \over dt} \right)_{\text{enhancement}}= {3\beta c \over R} n_\lambda 
\end{flalign}
We expect  equal numbers of photons of both helicities to emerge as decay products $n_+=n_-$, thus the total number density of photons $n_\gamma$ regardless of helicity $\lambda$ becomes $n_\gamma=2 n_\lambda$. Incorporating surface loss, density enhancement, and adding the photon number densities of both helicities, we adapt Eq.\eqref{d-lambda-dt} to an equation for $n_\gamma$,
\begin{flalign}\label{d-gamma-dt}
{d n_\gamma  \over dt} 
=&\Gamma_{a} \left[    n_a(t) 
[ 2 +   { 16\pi^2 n_\gamma(t) \over   m_a^3 \beta }  ]
  -  { 8\pi^2  [n_\gamma(t)]^2 \over m_a^3  }  (  {2 \beta\over3}  + 1 ) \right]
+ ( {3\beta c \over R} - {3c\over 2R} )   n_\gamma(t) ~.
\end{flalign}
The amplification of photon number density from decay of axion would at the same time deplete the number density of axions. This back reaction effect on the axion number density can be described by the following equation,
\begin{flalign}\label{da-dt}
{d n_a  \over dt} 
=&\Gamma_{a} \{  -   n_a(t) 
[ 1 +   { 8\pi^2 n_\gamma(t) \over   m_a^3 \beta }  ]
  +  { 8\pi^2  [n_\gamma(t)]^2 \over m_a^3  }  { \beta\over3}   \}
+ {3\beta c \over R}   n_a
\end{flalign}
which comprises terms on the RHS of Eq.\eqref{d-gamma-dt} except for a factor of $-1/2$. The third term on the RHS of Eq.\eqref{da-dt} is for back reaction of photons into axions from the same phase space of the initial axions \eqref{DistA}, but excluding the fast axions formed by energetic photons only. The production rate of fast axions $n_f$ is thus the difference between the last terms inside curly brackets of Eqs.\eqref{d-gamma-dt} and \eqref{da-dt},
\begin{flalign}\label{dfa-dt}
{d n_f  \over dt} 
={1\over2} \Gamma_a { 8\pi^2  [n_\gamma(t)]^2 \over m_a^3  }  
\end{flalign}
The equations for the dynamics of gravitational compression are
\begin{flalign}\label{dyn-eq1}
 - { G  m_a \over R^2} \left[ n_a(t)  {4\pi R^3\over3} m_a \right] =&  m_a {d^2 R\over dt^2} 
~,\qquad
\beta =  - {1\over c} {d R \over   dt  } ~.
\end{flalign}
The quantity inside the square bracket on the LHS of Eq.\eqref{dyn-eq1} is the total mass of the axion cluster, which changes as the decay process advances. The evolution of the axion-photon system is determined by coupled differential equations\eqref{d-gamma-dt}, \eqref{da-dt}, \eqref{dfa-dt}, and \eqref{dyn-eq1}. 

To summarize, this system of equations embodies spontaneous and stimulated decay of axions, the back reaction of photons into axions, the surface loss of photons from the region, the number density enhancement and the dynamics from gravitational compression. On the one hand, axion and photon number density change at a rate influenced by the number densities themselves, the speed of the axions, the radius of the axion cluster, and the lifetime of the axion. On the other hand, the radius of the axion cluster and the speed of axions are affected by axion number density and the radius itself through the dynamics of gravitational compression.

0000

Because QCD axions are weakly coupled to SM particles, the lifetime $\tau_a$ is so long that their decaying into photons are usually inefficient. Although stimulated decay improves the photon emission efficiency due to the Bose enhancement factor, the watershed moment of exponential amplification of photon number growth could still be postponed almost indefinitely if the axions cluster is simply not dense enough. Minimum and maximum axion cluster densities were estimated in stationary lasing scenarios\cite{Kephart:1994uy} as
\begin{flalign}\label{rhos}
\text{(min)}\rho_1={6\over K^4}({\text{eV}\over m_a})^2
~,~
\text{(max)}\rho_2=2({\text{AU}\over R})^2 ~.
\end{flalign}
By requiring the mean free path of photon to be shorter than the size of the cluster, or via the inequality $ 16\pi^2 n_a(t) /   m_a^3 \beta \tau_a > 3c / 2R $, meaning that photon production inside the cluster is faster than photon escaping off the surface, leads to $\rho_1$ which was taken as the minimum axion cluster for lasing to ignite. For stationary lasing, the condition $\rho>\rho_1$ is necessary but not sufficient, because an axion cluster typically requires the density $\rho$ to be meaningfully higher than $\rho_1$ in order to fully realize the potential of stimulated emission. The maximum density $\rho_2$ was obtained to keep the cluster from becoming a black hole. A large value of radius $R$ could make the $\rho_2$ small enough that $\rho_2\leq\rho_1$ which evicts all possible parameter space of cluster density $\rho$. Thus it was thought that for lasing to occur, the cluster radius $R$ could not surpass a maximum radius $R_1\sim{K^2\over2}({m_a\over\text{eV}})^2\text{AU}$. Note that a radius $R>R_1$ only stands for the relation $\rho_2\leq\rho_1$, and it is not a criterion to find out whether the cluster is a black hole or not. 

However, as will be demonstrated, one direct result from the gravitational compression is that, the allowed the parameter space of a lasing axion cluster to be expanded, and lasing can still happen even if the initial axion density $\rho_0 <\rho_1$ or the initial cluster radius $R_0 > R_1$. The reason is that in dynamical scenarios, despite the fact that the initial values of density and radius do not result in instantaneous lasing, the gravitational compression of the cluster evolves the initial parameter  to  critical values that trigger lasing. Hence, an axion cluster which was predicted to not lase using the stationary model can still lase when it is sufficiently gravitationally compressed.

\section{Results and examples}
We have solved equations \eqref{d-gamma-dt}, \eqref{da-dt}, \eqref{dfa-dt}, and \eqref{dyn-eq1} numerically for different values of system parameters (axion-photon coupling strength, axion mass, cluster mass, cluster radius and so on). Instead of arbitrarily selecting values for these parameters, we produced examples relatable or comparable to results presented in previous works of relevant topics. 

\subsection{Lasing in both stationary and dynamic scenarios}
Generally speaking, if an axion cluster of given size and radius is able to lase according to the stationary model, then it should also lase under gravitational compression. One criterion of lasing is high axion density, which only gets enhanced from compression.  
We examine three cases: the original static example, a cold example, and a warm example, each with the same initial mass and radius parameters.

\subsubsection{Stationary example}

\begin{figure}[ht]
        \centering
\includegraphics[width=0.65\textwidth]{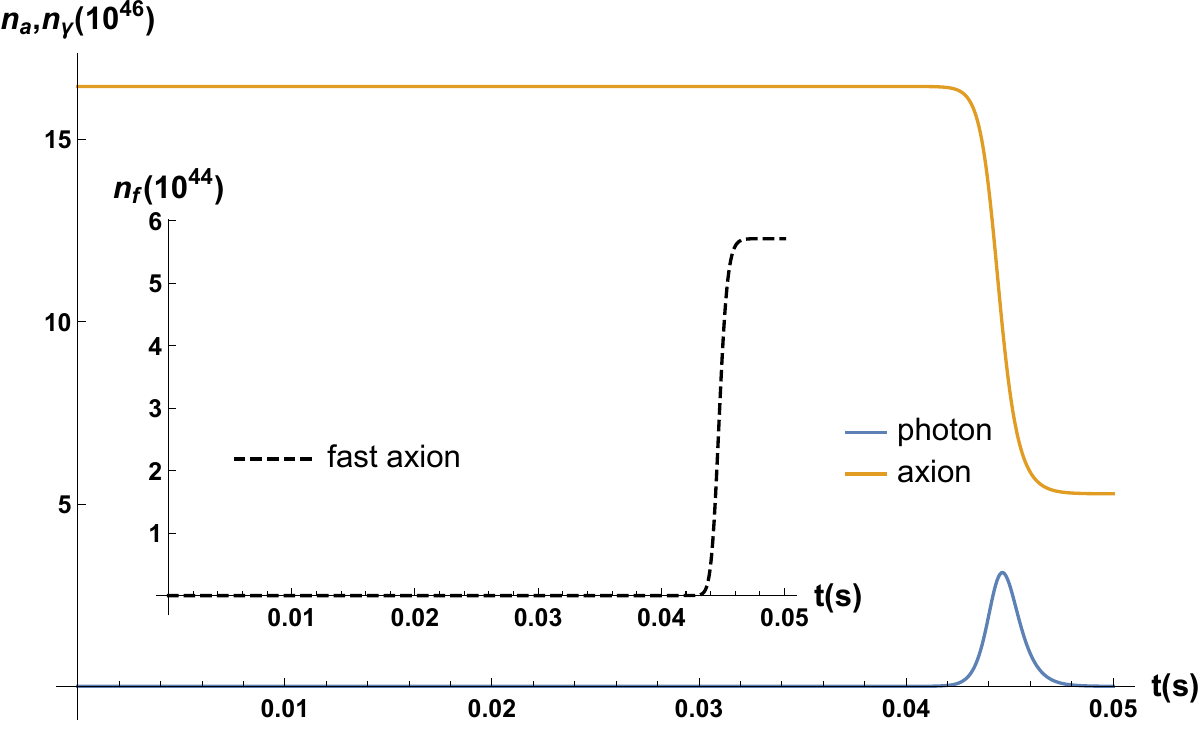}
\caption{\label{A1}
Number densities of axions, photons, and fast axions versus time during the stimulated emission from a cluster consisting of $m_a=10^{-4}$ eV axions with total mass $M_0=0.1 M_\odot$ and radius $R_0=1.8\times10^{-4} R_\odot$, based on the original stationary model.} 
\end{figure}

Consider first the stationary lasing of a cluster consisting of $m_a=10^{-4}$ eV axions with total mass $M_0=0.1 M_\odot$ and radius $R_0=1.8\times10^{-4} R_\odot$. The values of these parameters are not fine-tuned but are selected in such a way that they reproduce a working example (Example No. 3) from Ref.~\cite{Kephart:1994uy}.   The escape velocity of an axion from the self gravity of this cluster is $0.048c~(\beta=0.048)$ which is also set to be the maximum velocity of all axions in the cluster. Its initial density is $\rho_0=2.4\times10^{11}$ g/cm$^3$ with the minimum and maximum lasing density being $\rho_1=1.9\times10^9$ g/cm$^3$ and $\rho_2=2.8\times10^{12}$ g/cm$^3$, respectively. The maximum lasing radius $R_1=6.04\times10^{-3} R_\odot>R_0$, together with $\rho_1<\rho_0<\rho_2$, according to Ref.~\cite{Kephart:1994uy}, indicate that this cluster can lase, which is indeed the case as it produces a photon pulse within several tens of milliseconds. The evolving number densities of normal axions, photons and fast axions are shown in Fig.~\ref{A1}. 
At the beginning, the photon number is very low, hence it takes about 0.04 s to accumulate enough photons to trigger an explosive depletion of axions. Note that the peak value of photon number density $n_\gamma$ does not match the drawdown of photon number density $n_a$, owing to the fact that the photons are constantly escaping the cluster and only the excessive number of photons can form into a pulse. A small portion of energetic photons are converted into fast axions, which leaves the cluster region quickly. The number density of fast axions is about 1\% of that of the regular axions. After 0.05 s lasing stops because of the diluteness of the remaining axions. Photon productions from spontaneous and stimulated decay of axions are still ongoing but the newly produced photons mainly leave the cluster region before stimulating enough axions, which leads to no accumulation of photons. The end stage of the axion cluster after lasing was not discussed in the original stationary lasing model, but it was implied that the axion cluster would remain at a lower density state with negligible photon production.

\subsubsection{Cold dynamic example}

\begin{figure}[ht]
        \centering
\includegraphics[width=0.65\textwidth]{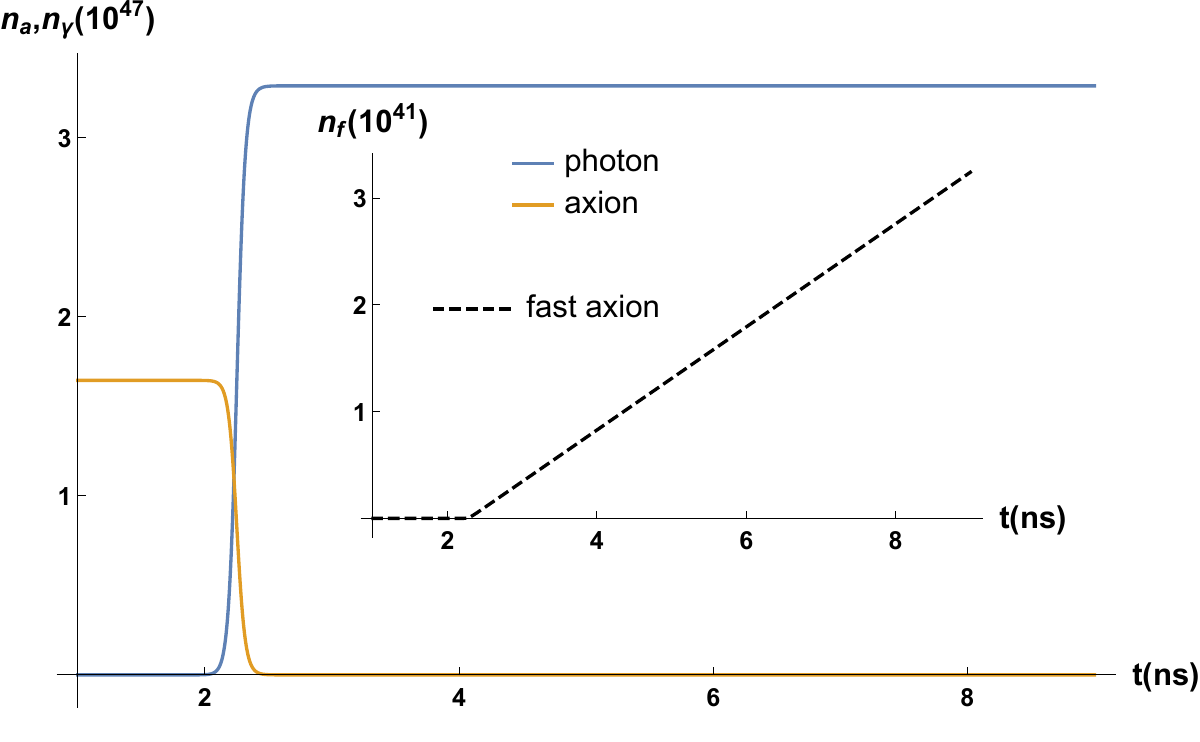}
\caption{\label{A2}
Number densities of axions, photons, and fast axions versus time during the stimulated emission from a cold cluster consisting of $m_a=10^{-4}$ eV axions with total mass $M_0=0.1 M_\odot$ and radius $R_0=1.8\times10^{-4} R_\odot$, based on the dynamical model.} 
\end{figure}

For dynamic lasing during gravitational compression, we again focus on  a cluster consisting of $m_a=10^{-4}$ eV axions with total mass $M_0=0.1 M_\odot$ and radius $R_0=1.8\times10^{-4} R_\odot$. In contrast to the stationary model, axions are assigned a initial velocity $\beta_0=v_0/c$ and subject to move according to Newtonian gravity. We consider  the case where all the axions are essentiallty ``frozen" in space with initial velocity $v_0=1$ m/s pointing towards the center of cluster. We plot the number densities of normal axions, photons and fast axions versus time in Fig.~\ref{A2}. 
These plots show that axions are quickly converted to photons in a fraction of ns, much faster than the stationary model in Fig.~\ref{A1}. (This somewhat surprising result is explained below.) The peak photon density is about twice that of axion, as the effect of photon loss from the surface is inefficient. The photons are still constantly leaving the surface of the cluster, but in a time measured by nanoseconds, the loss is negligible comparing to the burst of photon production.

The fast pace of the lasing process  depicted in Fig.~\ref{A2} is not a direct result of the dynamics, indeed, the process occurs so quickly even photons barely move. There was an aspect not discussed in the stationary model, but becomes a fundamental explanation for why the stimulated emission exhibits a rapid onset. The initial velocity $v_0=1$ m/s corresponds to a very temperature 
for $10^{-4}$ eV axions, classifying them as ultra-cold dark matter. The consequence of staying at this extreme low temperature is that the phase space of axions is being  crushed, making the occupation number extraordinarily high. A cold cluster has larger occupation numbers than those of a hotter cluster of the same density, which means cold axion clusters are more readily to lase; as the axions in cold clusters are closer to reaching Bose-Einstein condensation.

\begin{figure}[ht]
        \centering
\includegraphics[width=0.65\textwidth]{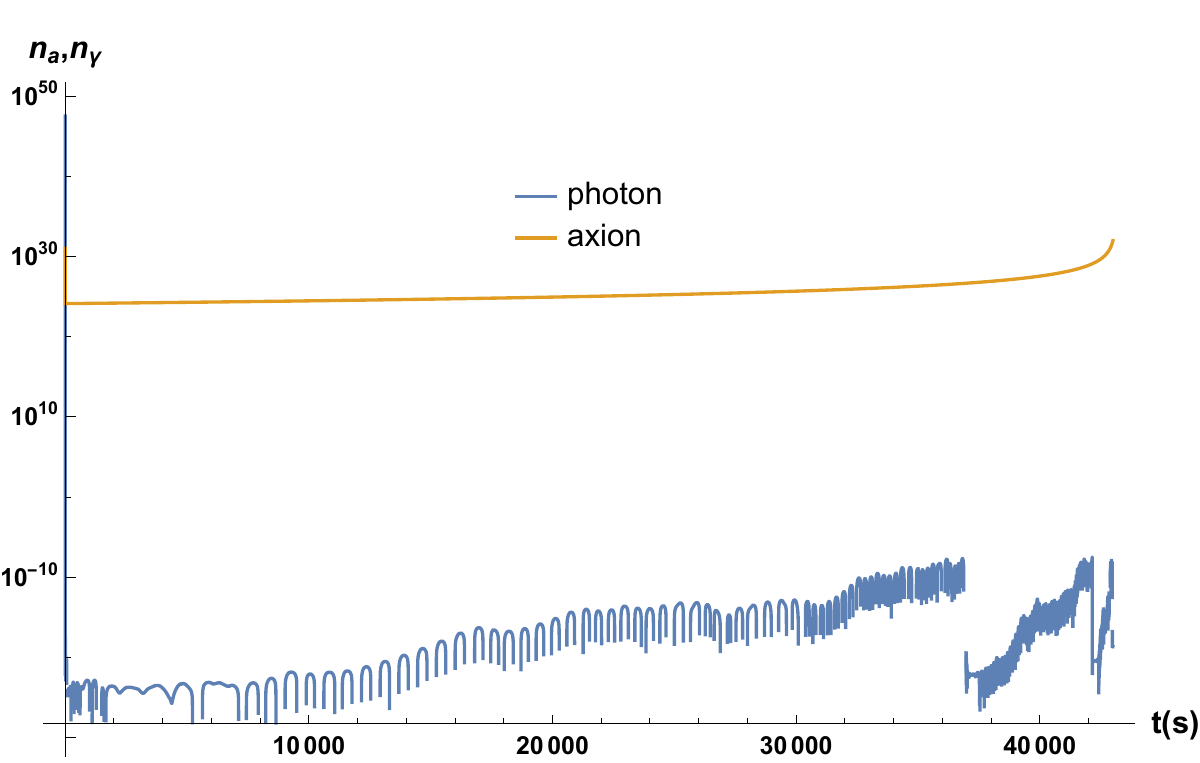}\\
\includegraphics[width=0.49\textwidth]{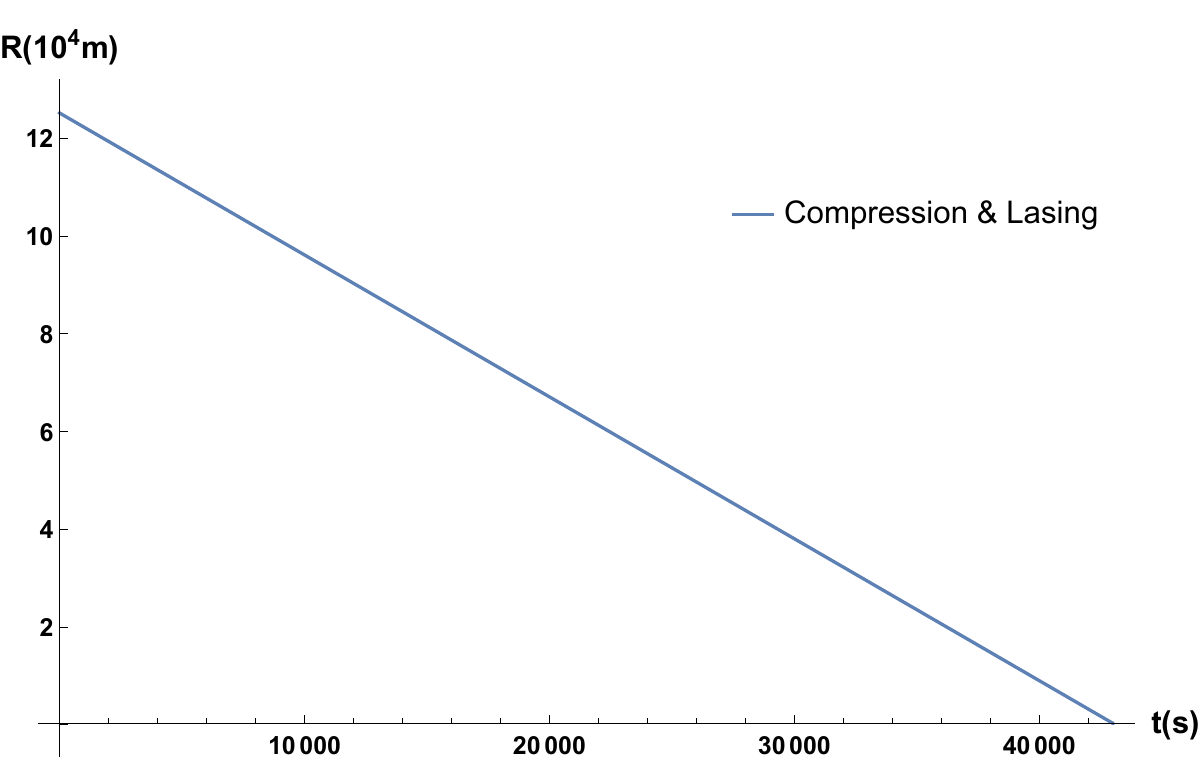}
\includegraphics[width=0.49\textwidth]{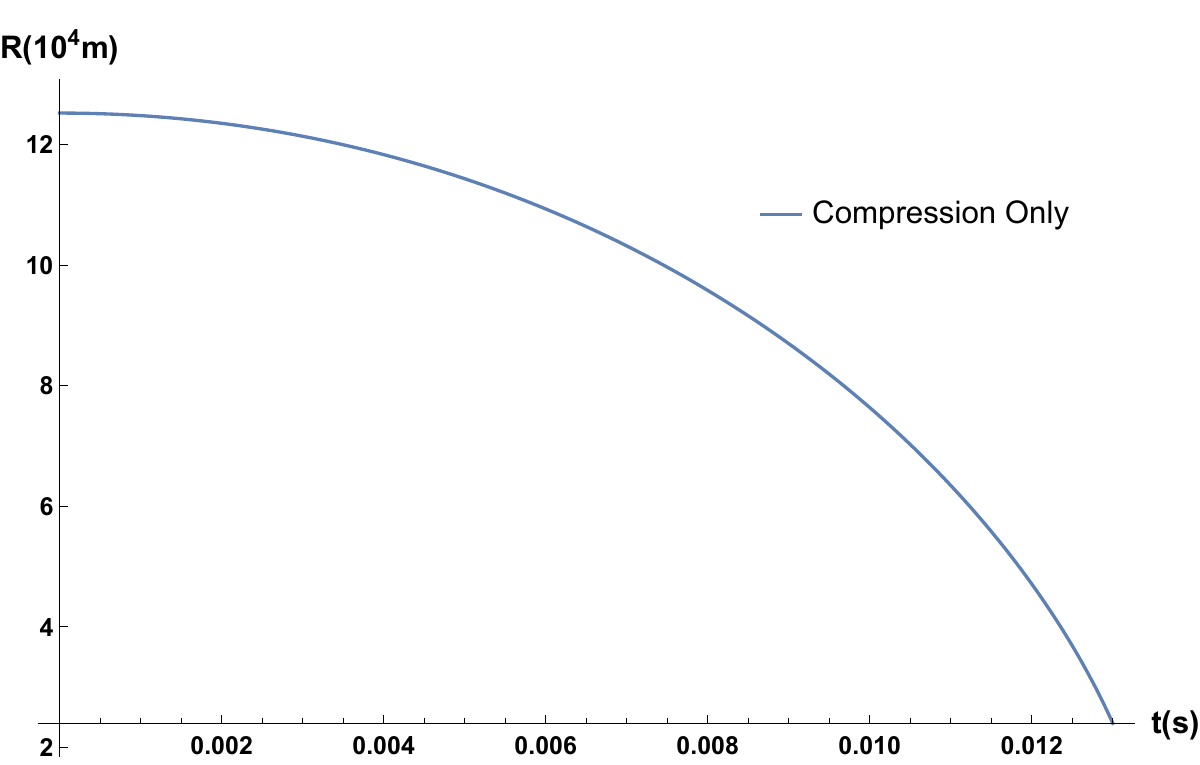}
\caption{\label{A3}
Number densities of axions and photons versus time after the stimulated emission from the cold cluster consisting of $m_a=10^{-4}$ eV axions with total mass $M_0=0.1 M_\odot$ and radius $R_0=1.8\times10^{-4} R_\odot$, based on the dynamical model(top). Reduction of the radius of the cluster from the combined effects of gravitational compression and stimulated emission of axions(Bottom left). Reduction of the radius of the cluster from the effect of gravitational compression only(Bottom right).}
\end{figure}

The dynamic lasing model offers the description of compression after the photon pulse formation. The top panel of Fig.~\ref{A3} shows the axion density continue to increases because of the ongoing contraction. In contrast the photon density is negligible  for the cluster being not dense enough to trigger another lasing event. The axion cluster lost so much mass from the stimulated emission in the first few nanosecond that it fundamentally changes the compression dynamics afterwards. The bottom left panel of Fig.~\ref{A3} gives the relation between the radius of axion cluster and time, showing that the cluster takes more than 14 hours before collapsing to a singularity. On the contrary, if one does not consider the lasing, a cluster of the same mass and radius would have collapsed to a singularity in about 13 ms, see bottom right panel of Fig.~\ref{A3}. The final estimated mass of the axion cluster is around $7.4\times10^{-25}M_\odot$, 24 orders of magnitude small than its original mass. Hence, if this axion cluster were to collapse into a primordial black hole, with the lasing process taken into account, the mass of the new born PBH would be much smaller.

\subsubsection{Warm dynamic example}

\begin{figure}[ht]
        \centering
\includegraphics[width=0.65\textwidth]{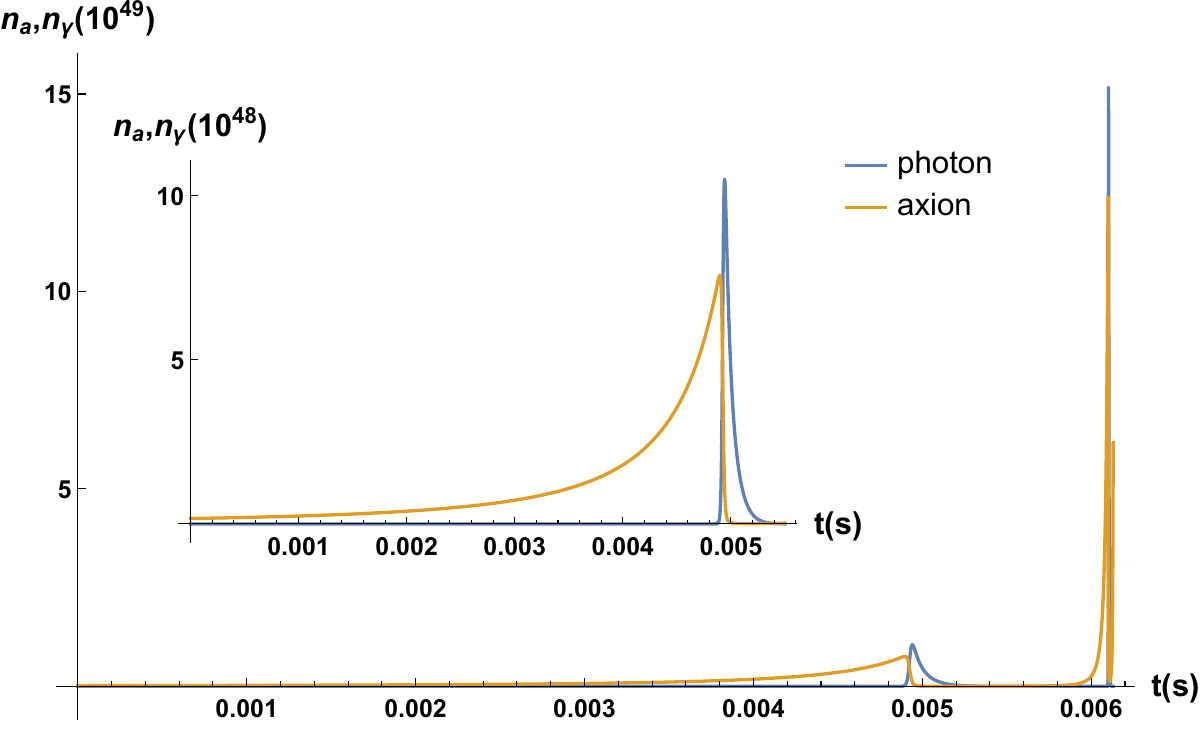}\\
\includegraphics[width=0.65\textwidth]{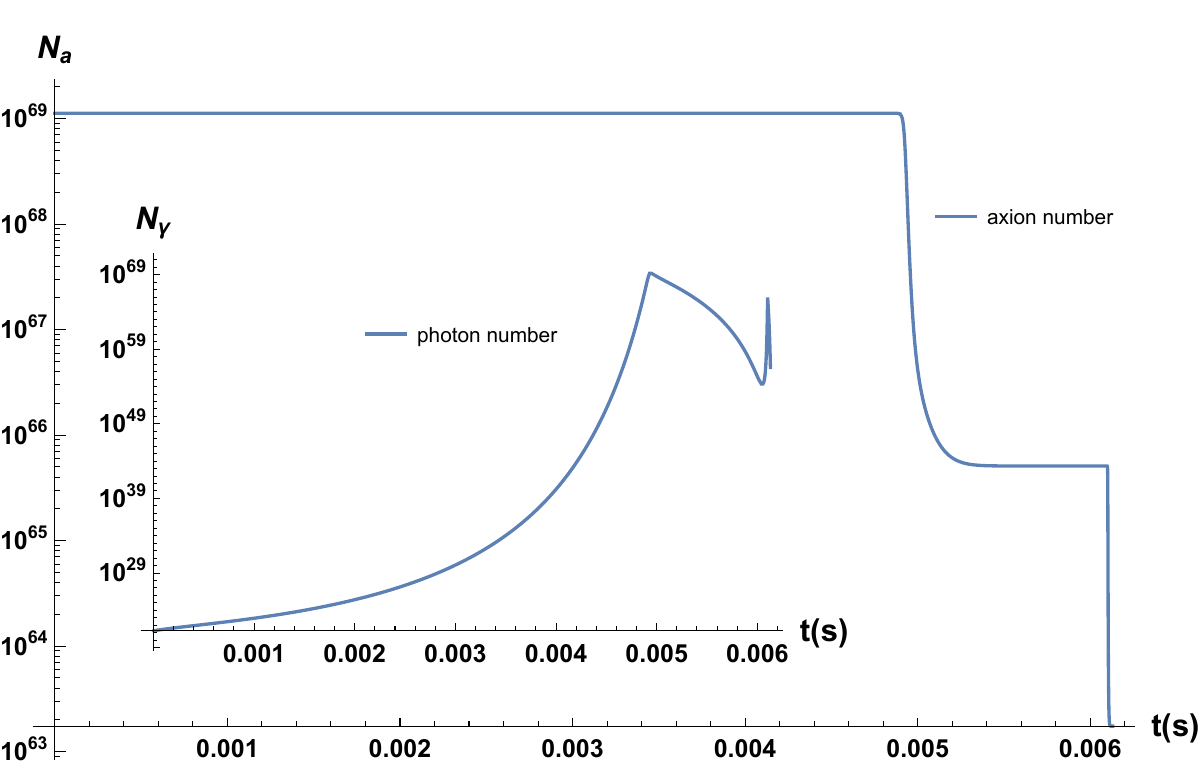}
\caption{\label{A4}
Number densities(top) and total numbers(bottom) of axions and photons versus time during the stimulated emission from a warm cluster consisting of $m_a=10^{-4}$ eV axions with total mass $M_0=0.1 M_\odot$ and radius $R_0=1.8\times10^{-4} R_\odot$, based on the dynamical model.} 
\end{figure}

Next, we look into a warmer cluster of $m_a=10^{-4}$ eV axions with the same initial mass and radius but the initial velocity being equal to the velocity used in the stationary model, i.e., $\beta_0=0.048$. The rates of change in axion and photon number densities are shown in Fig.~\ref{A4}.
 For comparison the top panel of Fig.~\ref{A4} present the evolving number densities at two time scales. The first lasing event occurs at around 5 ms with a peak photon number density greater than  $1.0\times10^{49}$, while the second lasing event takes place at around 6 ms with a peak photon number density reaching beyond $1.5\times10^{50}$.
Although the peak photon number density in the second pulse appears to be $\sim$15 times higher than in the first, but this comparison is misleading as it overlooks the fact that the initial lasing event is actually more intense. This intensity leads to the depletion of more axions, resulting in the production of a greater number of photons overall. The bottom panel Fig.~\ref{A4}  shows the two lasing events with the vertical axes representing the total numbers of axions and photons instead of number densities. 
In contrast to the ``frozen'' axion lasing event shown in Fig.~\ref{A2}, the initiation of lasing here is delayed by several orders of magnitude (nanoseconds vs. milliseconds). This is in line with the previous observation that colder axions are more easily stimulated to lase.

\begin{figure}[ht]
        \centering
\includegraphics[width=0.49\textwidth]{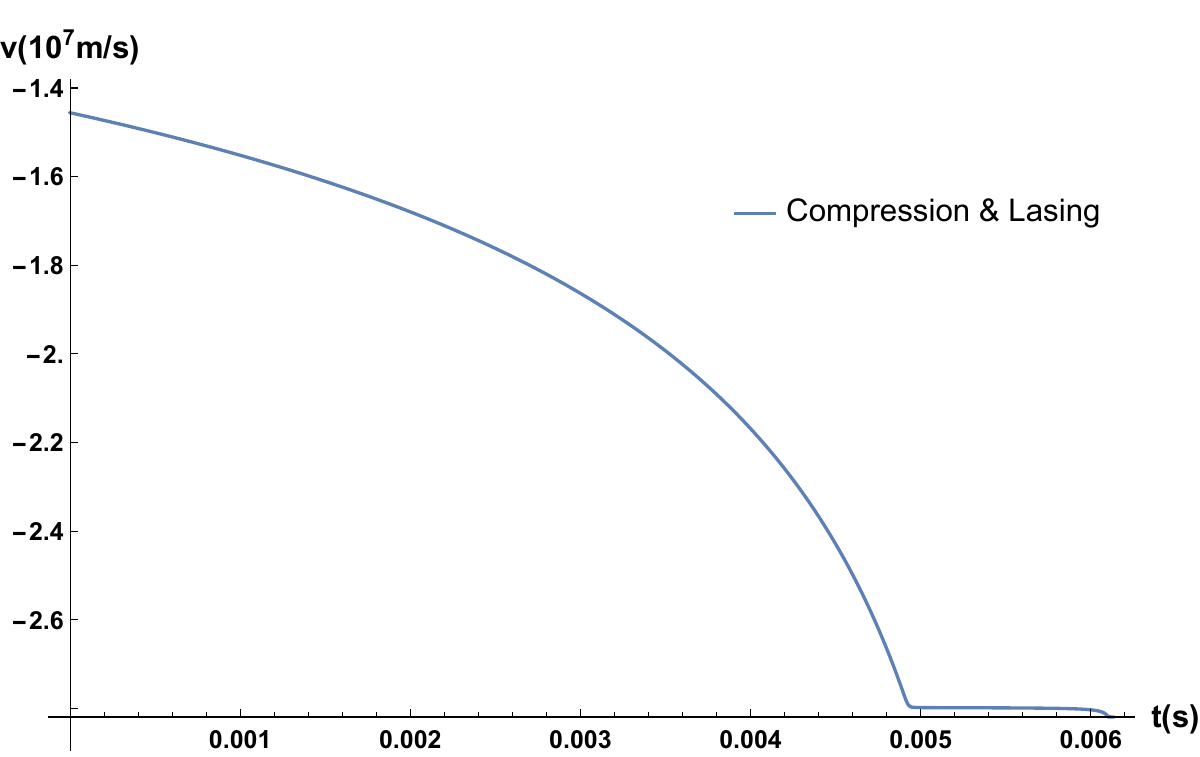}
\includegraphics[width=0.49\textwidth]{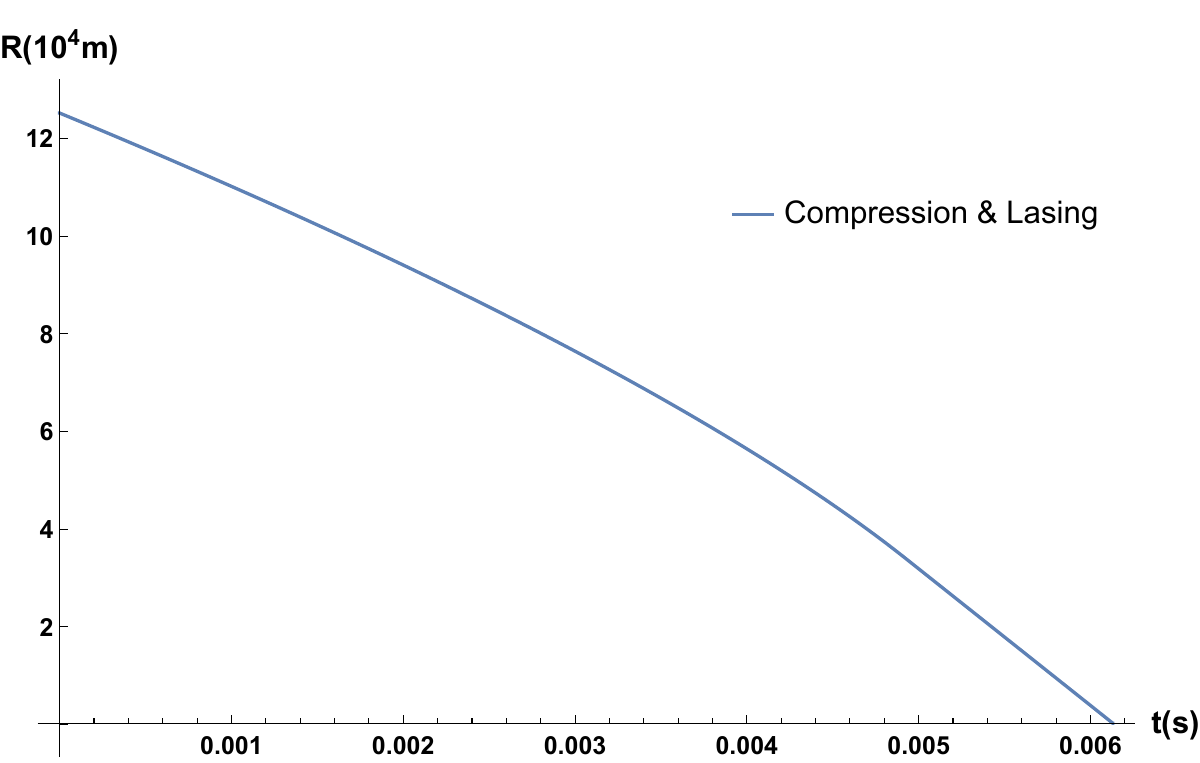}
\caption{\label{A5}
Velocity(left) and radius(right) of the warm cluster consisting of $m_a=10^{-4}$ eV axions with total mass $M_0=0.1 M_\odot$ and initial radius $R_0=1.8\times10^{-4} R_\odot$, affected by both gravitational compression and stimulated emission of axions, based on the dynamical model.} 
\end{figure}

The radius $R$ and the velocity $v=\dot R$ of the cluster's surface during compression in this scenario are depicted in Fig.~\ref{A5}. 
As can be seen from the left panel of Fig.~\ref{A5}, the first lasing event at around 5 ms is powerful enough to stop the growth in the velocity $\dot R$ due to gravitational pull because of the depletion of axion. The right panel of Fig.~\ref{A5} shows that after 6 ms the axion cluster continue collapsing towards a singularity, while the velocity of axions are still below 0.1$c$, which indicates that the non-relativistic treatment of the axions is appropriate. At the end of the time range in the figure, the mass and radius of the axion cluster are around $1.5\times10^{-7}M_\odot$ and 201 m respectively. The radius at the time is much larger than the Schwarzschild radius of $\sim$ 0.5 mm , evidently supporting the application of Newtonian gravity in this model. Here the mass of residue cluster is larger than that in the case of the ``frozen'' axions discussed previously, consistently indicating less efficient lasing for warmer axions. If a PBH were to be formed subsequently, its mass would come from the mass of this residue cluster.

\subsection{Lasing in only dynamic scenarios}

\begin{figure}[ht]
        \centering
\includegraphics[width=0.49\textwidth]{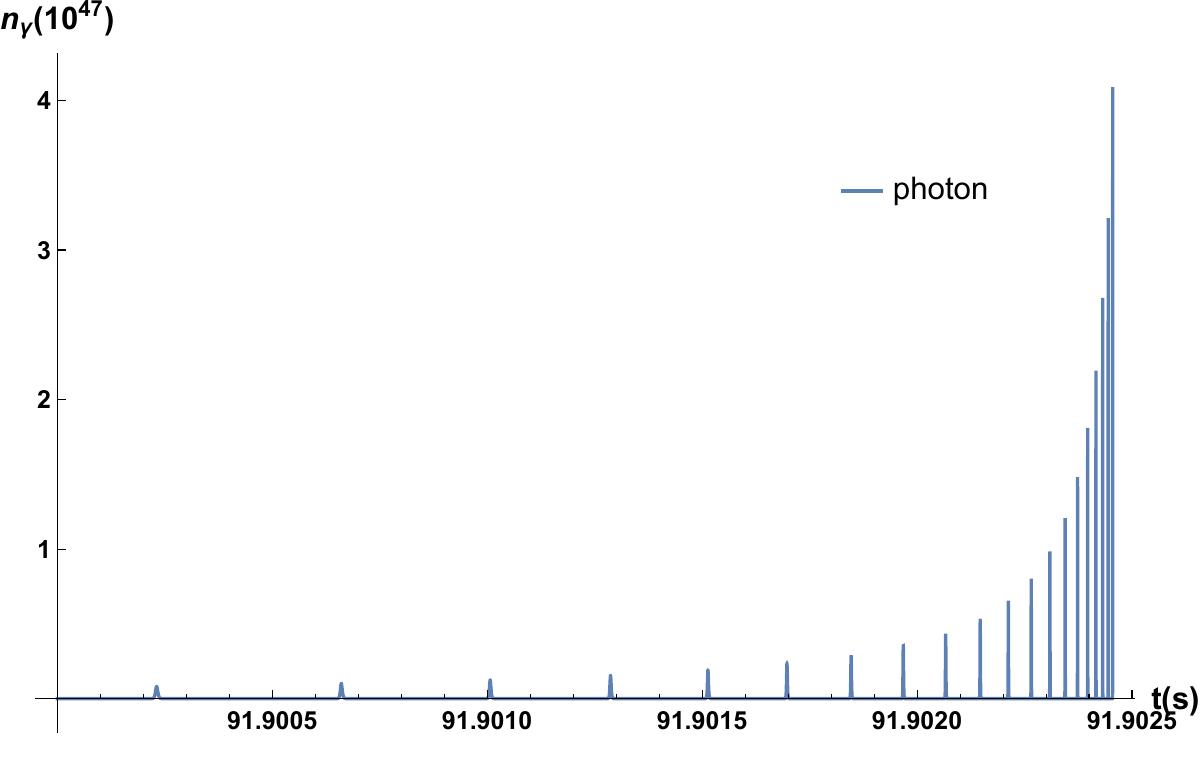}
\includegraphics[width=0.49\textwidth]{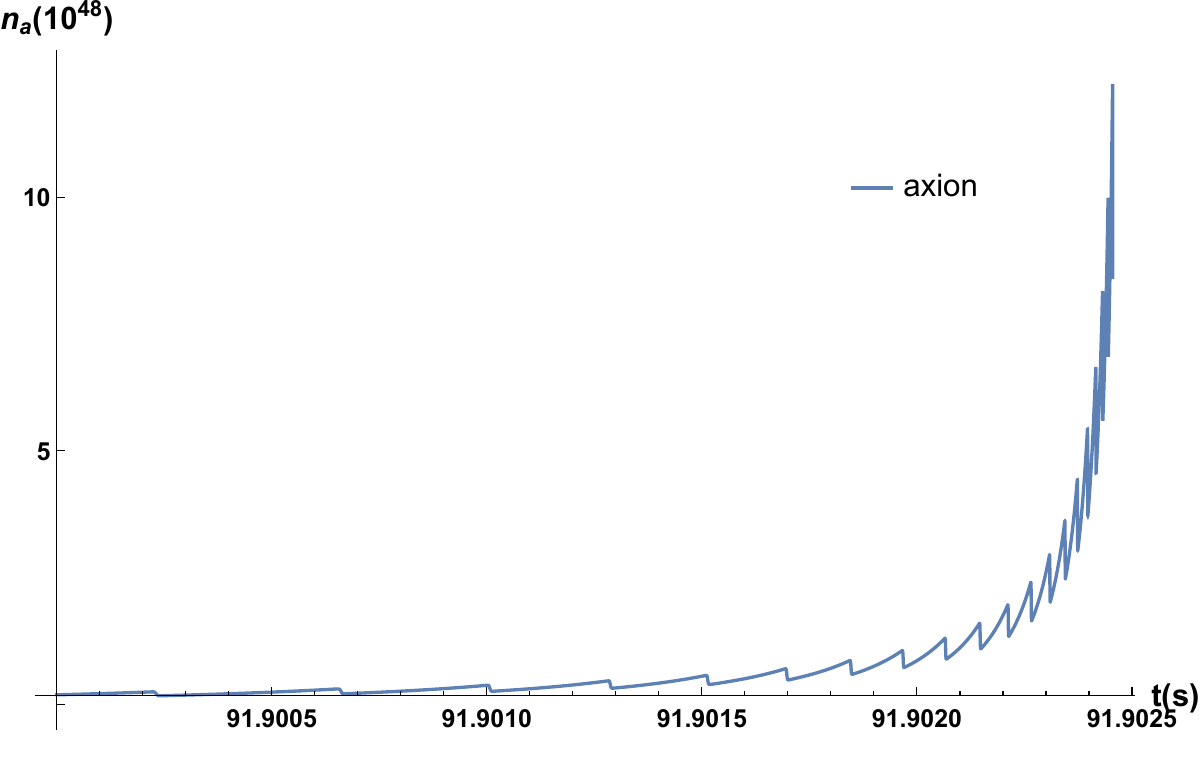}\\
\includegraphics[width=0.49\textwidth]{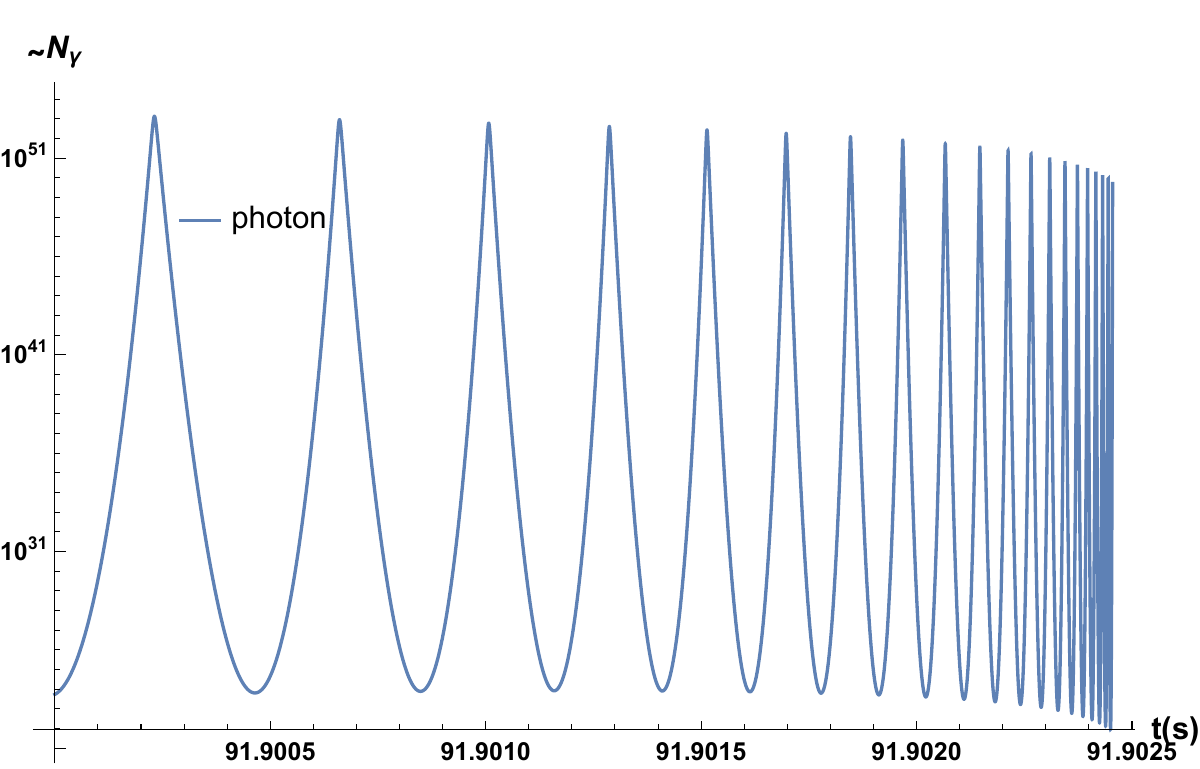}
\includegraphics[width=0.49\textwidth]{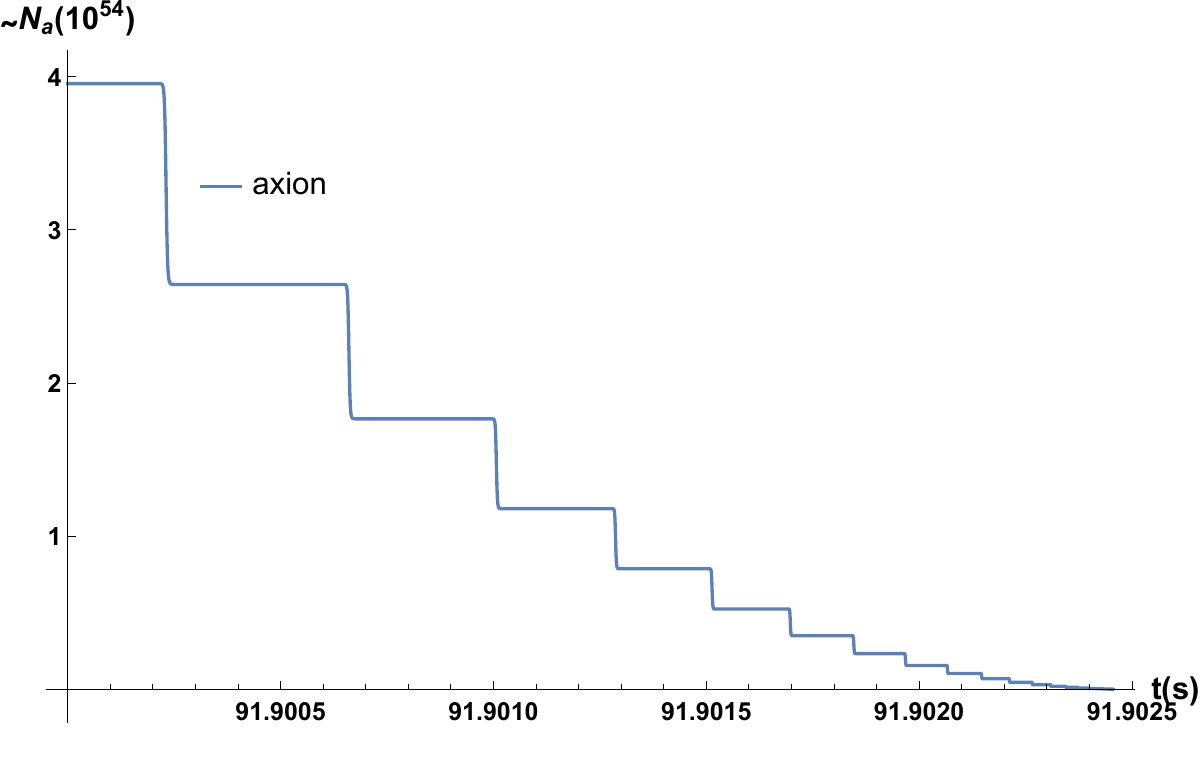}
\caption{\label{B1}
Number densities(top) and total numbers(bottom) of axions and photons versus time during the stimulated emission and the collapse of a cluster consisting of $m_a=10^{-4}$ eV axions with total mass $M_0=1.0 \times10^{-8}M_\odot$ and radius $R_0=3.0\times10^{-4} R_\odot$, based on the dynamical model. In stationary model, stimulated emission would not occur due to low initial axion density.} 
\end{figure}

We have shown examples where axions lase in both stationary and dynamic models. Moreover, even if an axion cluster of a specific size and radius fails to achieve lasing under the stationary model due to an insufficient axion occupation number, it may still be capable of lasing when subjected to gravitational compression. The compression would significantly enhance the axion occupation number, potentially enabling lasing to occur. The collapse of a self-gravitating Bose-Einstein condensate was studied by Chavanis in Refs.~\cite{Chavanis:2011zi,Chavanis:2016dab}. These studies highlight that axion clumps with masses exceeding a critical threshold $M_\text{c}$, can collapse into mini black holes quickly, where $M_\text{c}=1.012\hbar/\sqrt{G m_a |a_s|}$ and $a_s=-m_a/32\pi f_a^2$ is the scattering length. For QCD axions with mass $m_a=10^{-4}$ eV, $M_\text{c}=6.5\times10^{-14}M_\odot$, and $a_s=-5.8\times10^{-53}$ m. We demonstrate the lasing phenomenon resulting from the collapse of an axion clump with mass $M_0=1.0\times10^{-8}M_\odot\gg M_\text{c}$ and radius $R_0=3.0\times10^{-4} R_\odot$ in Fig.~\ref{B1} based on the dynamic lasing model. Note that stationary lasing model indicates that this clump will not lase as its density $3M_0/4\pi R_0^3$ is orders of magnitude below the minimum lasing density $\rho_1$.

In contrast, the dynamic clump not only has the ability to lase but also generates multiple pulses that exhibit a distinctive timing pattern. As the clump compresses, its density increases, prompting stimulated emission, with photons escaping from the clump quickly. This process subsequently repeats at an even faster pace due to the continuously decreasing radius of the clump. As the clump diminishes in size, photons escape more rapidly, resulting in a higher threshold axion density required to trigger stimulated emission, see the top right panel of Fig.~\ref{B1}. These increasing threshold axion densities correlate with elevated peak photon densities during the lasing process, as illustrated in the top left panel of Fig.~\ref{B1}. However, as shown in the bottom left panel of Fig.~\ref{B1}, the peak number of photons produced actually decreases, as the increase in density cannot offset the reduction in clump size. The total number of axions decreases in a stepwise manner as lasing events occur, which is illustrated in the bottom right panel of Fig.~\ref{B1}.

In this case, the dynamics of gravitational compression remain largely unchanged by lasing events for most of the evolution, as stimulated emissions primarily occur during the final stages of the collapse. The remaining mass of the clump at the end of the timeframe in our calculation is approximately $8.9\times10^{-14} M_\odot\ll M_0$. If the end product of the collapse were a black hole, its mass would be a small fraction of the mass it started the collapse.

\begin{figure}[ht]
        \centering
\includegraphics[width=0.49\textwidth]{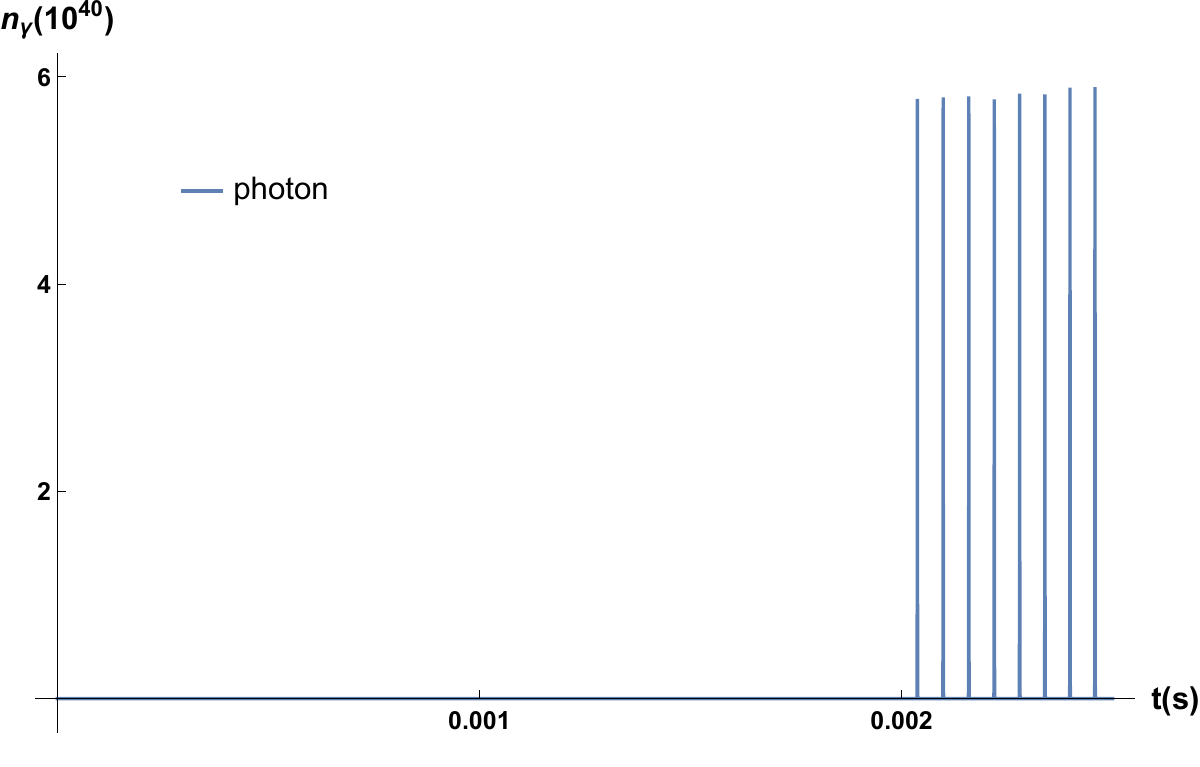}
\includegraphics[width=0.49\textwidth]{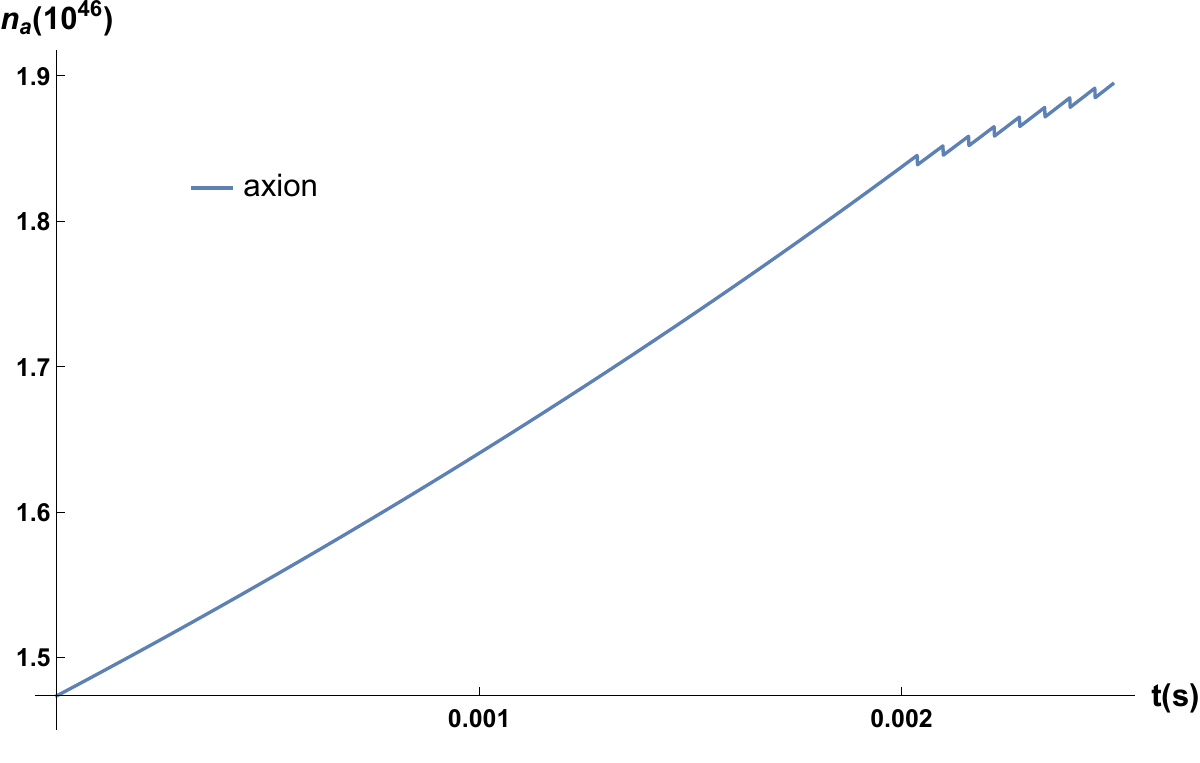}\\
\includegraphics[width=0.49\textwidth]{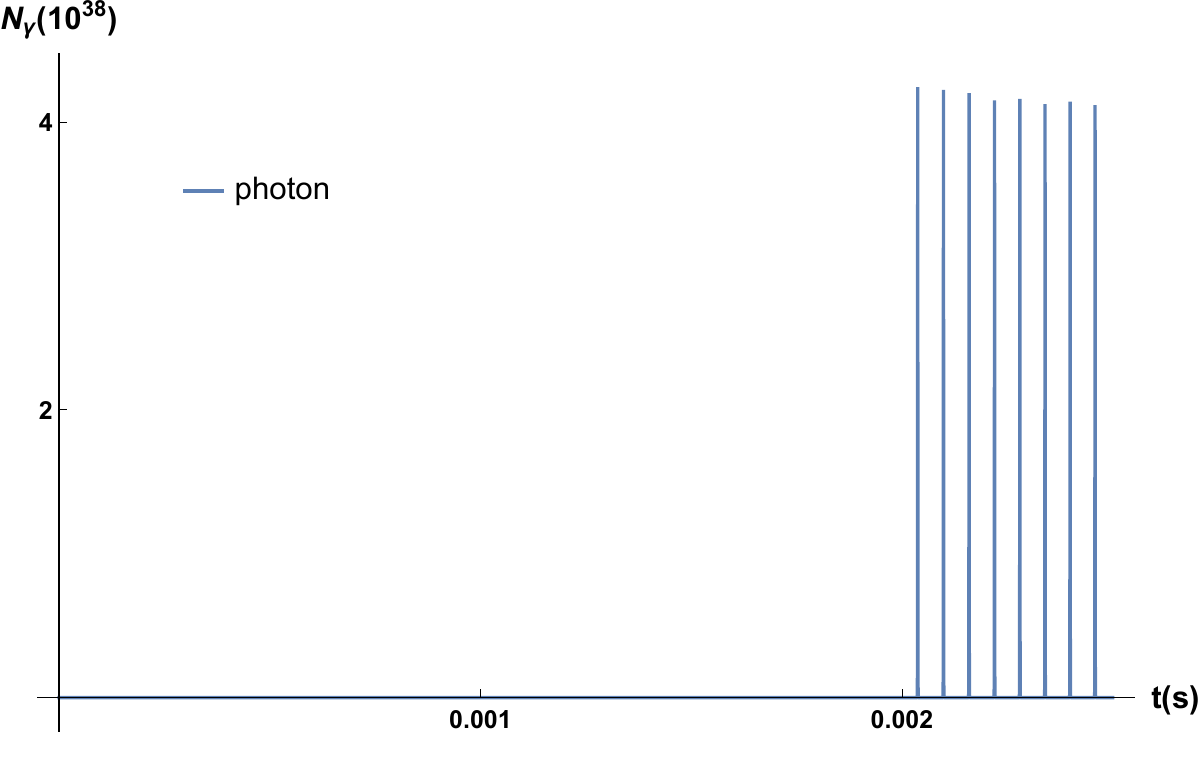}
\includegraphics[width=0.49\textwidth]{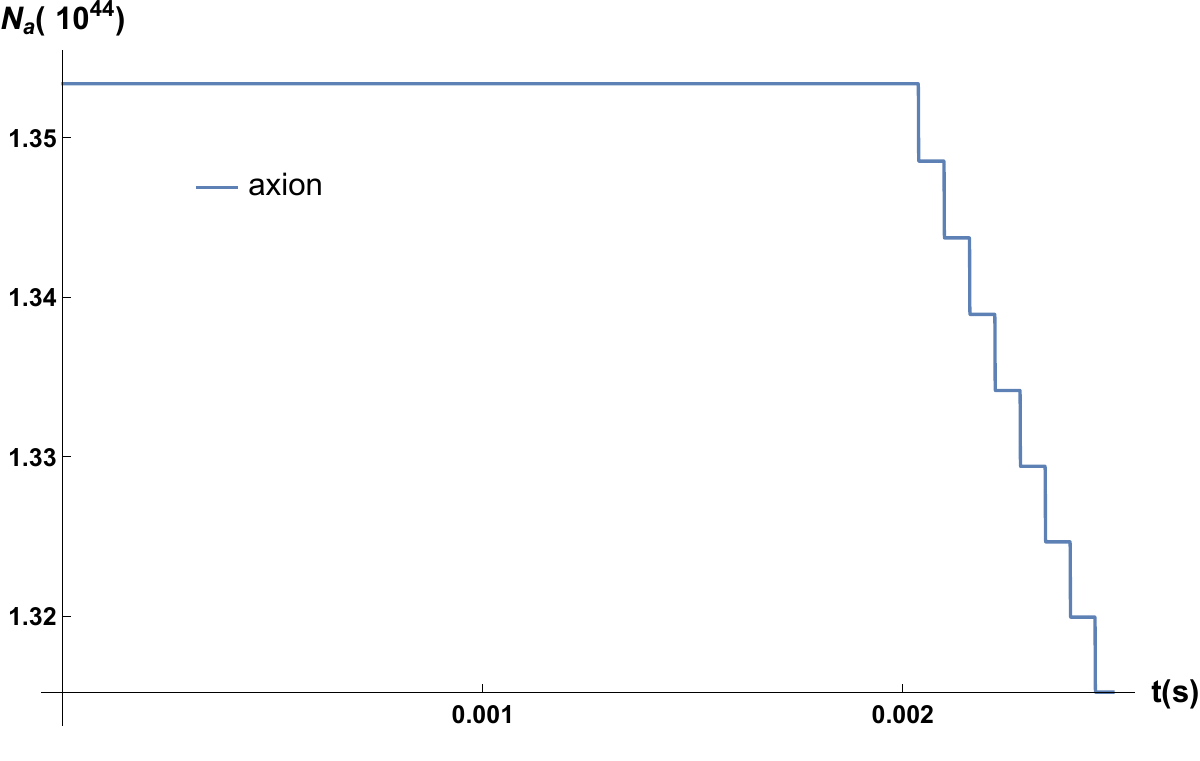}
\caption{\label{B2}
Number densities(top) and total numbers(bottom) of axions and photons versus time during the stimulated emission and the collapse of a cluster consisting of $m_a=10^{-4}$ eV axions with total mass $M_0=1.0 \times10^{-20}M_\odot$ and radius $R_0=1.0\times10^{-6} R_\odot$, based on the dynamical model. In stationary model, stimulated emission would not occur due to low initial axion density.} 
\end{figure}

For a dilute axion cluster, lasing can still occur with enough time of gravitational compression. Let consider an axion cluster with initial mass $M_0=1.0\times10^{-20}M_\odot\ll M_\text{c}$, and radius $R_0=1.0\times10^{-6} R_\odot$, which give a very low density of 14 kg/m$^3$. The varying number(densities) of photons and axions during late stage of the collapse are shown on the bottom(top) panel of Fig.~\ref{B2}.
It takes more than 2 hours for the axion cluster to be compressed and to initiate the lasing process, with an initial velocity of 0.06 m/s. The remaining mass of the initial axion cluster after releasing photon pulses is around $9.7\times10^{-21} M_\odot$, indicating that only 3\% of axion mass is lost from stimulated emission. The initial mass of the cluster is low so that it needs to collapse to a much smaller radius for stimulated emission to be triggered. However, this significantly shortened radius is prone to photon escaping which results in occurrences of stimulated emission being short lived and thus ineffective axion consumption. The dynamics of the collapsing process is not impacted by lasing events in this example since mass loss from stimulated emission is small.

\section{Discussions}

In Sec.~\ref{sec-intro} we outlined the reasons  why stable clumps are not ideal objects for investigating the lasing phenomenon. These stable clumps have not been well classified, as tests for their photonic response have not been conducted. The photonic response is likely to alter the clump configuration significantly, which contradicts the purpose of using stable clumps as the research focus. Additionally, studies on axions often rely on Gaussian\cite{Chavanis:2011zi,Chavanis:2016dab} or sech\cite{Hertzberg:2018zte} ansatze, which may presuppose specific clump configurations of the forms
\begin{flalign}
\label{ansatzes}
\psi({\bf r},t)=\left \lbrack \frac{M}{\pi^{3/2}R(t)^3}\right
\rbrack^{1/2}e^{-\frac{r^2}{2R(t)^2}}e^{imH(t)r^2/2\hbar}, \,\,\,\, \text{or\,} 
~~~~
\psi(r) =\sqrt{3\,N\over\pi^3\, R^3}\mbox{sech}(r/R)
~.
\end{flalign}
These configurations may not properly represent initial density profiles as they imply that the cores of the clumps are much heavier than the edges. This scenario usually occurs after certain processes, such as gravitational compression, have been completed. If one were to adopt these density ansatze \eqref{ansatzes} in investigations of the photonic resonance of axions, it could unintentionally imply that resonance occurs after gravitational compression. In contrast, our analysis begins with a uniform density distribution and incorporates both mechanisms within the same time frame, providing a more unified description of the process.

Depending on the axion mass, the photons from its decay would fall within the radio-microwave range. Compared to other radio signals, such as those from pulsars and quasars, the least understood radio phenomena originating at extra galactic distances  are perhaps the fast radio bursts(FRBs, see Ref.~\cite{Lorimer:2024ysl} for a recent review). Due to a lack of evidence for celestial axions, researchers have explored the possibility that the rapid annihilation of axions into photons might be a mechanism for producing FRBs. 
Several astrophysical scenarios suggest that axion lasing may be possible, including from dense dark matter axion clusters. Ref.~\cite{Rosa:2017ury} illustrated that the stimulated(quantum) emission of axions produced by black hole superradiance can generate repeating laser bursts and pioneered the idea that these bursts are possibly linked to the observed FRBs. It was later confirmed that the bursts of EM radiation from superradiant axion clouds can also occur at the classical level\cite{Ikeda:2018nhb}. 

In addition to the superradiant axion clouds, we propose that the dynamic lasing of axion stars could serve as another candidate source of (repeating) FRBs through gravitational compression and stimulated emission. Depending on parameters such as the radius and mass of an axion star, the photonic bursts may occur as either repeaters or one-time events. For superradiant axions, the axion-black hole coupling $G m_a M_\text{BH}/(\hbar c)$ needs to be small $\lesssim1$ which only allows primordial black holes to produce superradiant QCD axions. Moreover, these primordial black holes must possess a high spin, as superradiance extracts the rotational energy of a black hole, which is typically achieved only through merger events. In contrast, the dynamic lasing of QCD axions is not subject to the restrictions pertaining to superradiant axions, and does not require a Kerr primordial black hole at all.

Furthermore, since the photonic pulses from superradiant axions are typically produced at a steady rate \cite{Rosa:2017ury,Chen:2023bne,Spieksma:2023vwl}, a quenching mechanism is necessary to explain why FRB repeaters are detected only intermittently. Ref.~\cite{Rosa:2017ury} suggests that this quenching mechanism arises from plasma generated by Schwinger pair production; however, this view is not universally agreed upon, as Ref.~\cite{Spieksma:2023vwl} found that plasma would not quench the instability. In contrast, the production of bursts from the dynamic lasing of axion stars does not require quenching, as it would automatically cease when gravitational compression concludes and the stars collapse. Therefore, the dynamic lasing of axion stars is a more plausible explanation for the transient nature of some FRB repeaters.

There have been a varity of results in the literature about the remnants  of collapsing axions clumps. Refs.~\cite{Chavanis:2011zi,Chavanis:2016dab} found that axion clumps with mass beyond $M_c$ are expected to collapse into mini black holes, while the collapse stabilizes and no black hole is formed according to Ref.~\cite{Eby:2016cnq}. Ref.~\cite{Guerra:2019srj} even suggests a middle ground, that unstable axion stars either collapse into black holes or migrate to a stable configuration.
In this work, regardless of whether the remnant objects of collapsing axion clouds are black holes, we estimated the masses of these remnant objects and found results that were neglected in previous studies on the collapse of axions. First, stimulated emission plays an important role by converting most axions into photons that escape, resulting in the remnant mass being only a small fraction of the initial clump mass in some cases. Second, stimulated emission prolongs the gravitational compression process, as it constantly reduces the clump mass.

\section{Conclusions}
Axions are a natural choice for the dark matter component of the standard $\Lambda$CDM model of cosmology. Axions are produced at rest in the QCD phase transition. Quantum fluctuations of the axion field that undergo inflation and are pushed outside the horizon return as scale-invariant density perturbations. We expect axions to be clumped on all scales and take this as motivation to study these configurations. A simple approximation is a uniform spherically symmetric clump of axions with vacuum outside. In strong gravity regime, the compression of the clump can be described by pressureless spherical gravitational collapse, an Oppenheimer-Snyder solution to the Einstein equations. Here we have adopted a non-relativistic treatment, as Newtonian gravity is applicable and the terminal speeds of axions are non-relativistic.
Previous works\cite{Hertzberg:2018zte,Levkov:2020txo} on this topic show that parametric resonance is not possible for QCD axions, even when considering collapsing axion stars. However, these studies primarily focused on stable axion clumps and drew their conclusions based on that assumption. In this work, we do not assume that the axion clumps are in a ``stable'' configuration. Instead, we find that the combined effects of gravitational compression and stimulated emission from axion clumps can produce repeating photonic bursts, offering an alternative explanation for the observations of some FRB repeaters, even in the context of QCD axions.

Some axion configurations have decayed through lasing and alter the composition of density perturbation in the early Universe. This in turn should have changed the spectral index away from purely scale invariance and could be compared with observation. Such an analysis would be interesting to carry out, but is beyond the scope of the present work.

\bibliography{draft01}

\end{document}